\documentclass[prl,aps,showpacs]{revtex4}
\usepackage{ifpdf}
\usepackage{amsmath}
\usepackage{graphicx}
\usepackage{color}
	\usepackage{amssymb}

\begin{document}

\title{Do microscopic stable black holes contribute to dark matter?}
\author{P. Suranyi, C. Vaz and L.C.R. Wijewardhana}
\affiliation{Department of Physics, University of Cincinnati, Cincinnati, Ohio, 45221-0011}

\begin{abstract} 
We investigate some of the experimental, observational and theoretical consequences of 
hypothetical stable black holes in the mass range between the electro-weak scale and the 
Planck mass, 2.4$\times 10^{15}$ TeV.  For the purpose of calculations we use Lovelock 
black holes in odd dimensions. If such black holes exist they contribute to dark matter. 
We show that the passage of the black holes through matter and the collision of black 
holes have a well defined experimental signature.  Depending on their cross section and 
energy they also accumulate in stars and influence their development. 
 \end{abstract}
 \maketitle
\section{Introduction}

Gravity theories in space-time dimensions greater than four have been studied extensively ever since the seminal
papers  of Kaluza and Klein on unification of gravity and gauge forces were published in the early twentieth 
century. Modern attempts at unification of forces at the quantum level, supergravity and super string theory are 
formulated in eleven and ten dimensions respectively.  Low scale gravity models proposed by  Arkani  Hamed,  
Dimopolous and Dvali~\cite{ADD} or Randall  and Sundrum ~\cite{RS1}~\cite{RS2}   to explain the hierarchy 
problem are also formulated in dimensions greater than four. In all these models bulk extra dimensions have to be
compactified or made small by warping to explain why the observed universe has only four space-time dimensions.

These extra dimensional gravity models admit black hole or black brane solutions. Masses of these black objects 
could range from the fundamental Planck scale of the theory up to astrophysical mass scales of millions of solar 
masses. Microscopic black holes in these models would have sizes smaller than the scale of compactification while 
large astrophysical black objects would more likely be black strings or branes.   Planck scales in these models  
could vary  from  standard $M_P=2.4 \times10^{15} $ TeV down to 1 TeV, in low scale gravity models. In particular, 
low scale gravity models admit microscopic black holes of TeV mass.
Such objects could be produced in particle accelerators like the LHC \cite{dimopoulos} or in collisions of high 
energy cosmic rays with matter \cite{cosmic}. Most of the studies investigating black hole production at the LHC 
assume that the produced black holes are of the standard Schwarzschild-Tangherlini type and will decay almost 
instantaneously into elementary particles via Hawking radiation~\cite{production}.  An exception is the recent 
work of Giddings and Mangano \cite{giddings}, which examines the consequences of the very conservative 
assumption that Hawking radiation does not exist at all and conclude that TeV scale black holes produced at 
the LHC would pose no risk to the Earth over its lifetime. 

In this paper we will investigate the fate of the black hole relics in a model, which incorporates standard 
Hawking radiation but in which black hole masses have a lower bound.  Upon approaching the mass bound 
the Hawking temperature of these black holes goes to zero and they become stable in vacuum.  Their cross 
section for collisions with elementary particles and each other would, however, stay finite, leading to 
observable consequences when they come into contact with matter or collide with each other inside stars 
or in the galaxy.  

If there are  stable microscopic primordial black holes then one has to confront the possibility that they could 
have a relic density large enough to over close the universe.  For this to become a serious  issue the temperature 
of the universe at some epoch has to be comparable to the  minimum stable  mass of the black holes  so that 
they could be produced in particle collisions. For the current discussion  we assume that  the reheating 
temperature of the universe after inflation is always much  less than the minimum mass of the black holes.   
We plan to  discuss the  issue of the relic density of stable Lovelock black hole remnants  in another publication.

According to our hypothesis all primordial black holes, or black holes produced by cosmic rays, that would 
otherwise have decayed by the present epoch should end up in (or very close to) the minimal mass end state.  
Such black holes contribute to or may even dominate dark matter.   Their number can only decrease in 
collisions with  each other or by absorption by macroscopic black holes. We will discuss the passage of 
stable black holes through matter and point out that their interaction with matter would have a characteristic 
signature, detectable in underground experiments. 

 Lovelock gravity theories admit black hole solutions with a limiting minimum mass and vanishing Hawking 
temperature. There is no reason known to us why in dimensions higher than four the contribution of Lovelock 
terms \cite{lovelock}, formed from higher powers of the curvature tensor, to the Einstein action should be 
suppressed.  Now in odd dimensions the ADM masses of black holes are limited below by the coupling strength 
of Lovelock terms. Furthermore, when such black holes evaporate by Hawking radiation their Hawking 
temperature approaches zero when their ADM mass approaches the lower bound.  The prime example of 
this is black holes in 5-d Einstein Gauss-Bonnet gravity.  Therefore, such a scenario is not completely in the 
realm of phantasy. 

Since the world has 4 non-compact dimensions on macroscopic scales, we need to compactify all but four 
of the odd number of dimensions in which the stable minimum mass black holes live. If the size of a black 
hole is smaller than the compactification scale, then the caged black hole should still have a stable minimum 
mass. A systematic expansion in the inverse of the compactification scale for black holes in Einstein gravity 
has been studied in the past \cite{harmark} \cite{karasik}.  Lovelock black holes can be embedded into a 
higher dimensional space, turning the black holes into black branes in those extra dimensions.  While we 
return to the  discussion of these questions in the last section of this paper we will study these and similar 
problems more extensively in a future publication. 

A simpler alternative, of trying to compactify all dimensions beyond 4 at the same scale in Gauss-Bonnet 
gravity has recently been attempted \cite{blackbrane}. Though a minimal mass exists in such a scenario, 
but the Hawking temperature does  not vanish and consequently black branes are not stable at the 
minimal mass.

The consequences of Gauss-Bonnet gravity on black hole phenomenology have already been considered 
by Rizzo~\cite{rizzo1}~\cite{rizzo} concentrating on the signature of TeV-scale stable black holes produced
at LHC, and by Alexeyev, Barrau, Boudoul, Khovanskaya, and Sahzin \cite{alexeyev} in the framework of 
an ADD \cite{ADD} or Randall-Sundrum II \cite{RS2} model with limiting masses of macroscopic range 
(1 $g$- $10^{18}\, g)$. We would like to concentrate on models in which the minimal mass of black holes 
is comparable or below the 4 dimensional Planck mass.  Such a mass limit would be natural for example 
in the framework of a Randall-Sundrum \cite{RS1} scenario, ADD theory in 9 or 11 dimensions, or Kaluza-Klein 
models in 11 dimensions and some string-inspired models of compactification. 

Though Schwarzschild-Lovelock black holes in odd dimensions are at the back of our minds, our investigations 
are fairly independent of specific models. They require the knowledge of a few basic parameters, namely  
the coupling strength of the Lovelock terms, to describe the near-minimal mass behavior of these solutions. 
We assume that all non-topological Lovelock terms are present in the Lagrangian and all of their couplings 
are in the scale of the Planck mass.  It is worth mentioning that higher order corrections to heterotic string 
theories contribute a Gauss-Bonnet term (second order Lovelock term) to the classical action \cite{heterotic}, 
\cite{boulware}. 

When all the Lovelock coefficients are of $O(1)$, black holes or black branes with $M>>M_P$  are well 
described by the Schwarzschild-Tangherlini  metric, their behavior is quite standard and well described by 
Einstein gravity.  In this paper we will concentrate on black holes with masses  close to the minimum mass 
of the order of $M_P$, when Lovelock terms play an important role leading to significant 
phenomenological consequences.

In the current paper we have not addressed questions pertaining to detection of stable microscopic black 
hole remnants by direct dark matter detection experiments \cite{dama}, \cite{cdms}. On one hand, if 
the minimum mass of these black holes is close to 1 TeV, then we will show that their abundance must be 
extremely low compared to other components of dark matter to avoid a contradiction with proton decay 
experiments.  On the other hand, if their mass is much higher it could be difficult to detect them in direct 
dark matter detection experiments. 

Finally, to close this section we list the issues  we intend to address in the rest of this paper. (i) Stable minimal 
mass black holes aggregate in clumps under the influence of gravity. Then they will form part (or all) of 
dark matter. This effect was considered for macroscopic stable black holes by Alexeyev et. al. \cite{alexeyev}, 
and for even larger, astronomical size black holes by Frampton \cite{frampton}, as well. These black holes 
may accumulate inside stars depending on the parton cross section, their velocity distribution, and the 
gravity at the surface of the star.   We will discuss the rate of heat production by the collision of accumulated 
black holes and compare it  to that of "standard" WIMPs \cite{kouvaris}.  (ii) Black holes, as part of dark 
matter, would collide with partons while passing through detectors in underground laboratories. These 
collisions would be followed by a characteristic decay that is quite different from those of Schwarzschild 
black holes.   Using past proton decay experiments we are able to give a rough lower bound on the 
minimum mass of stable black holes and/or their abundance compared to other dark matter constitutents. 
We also propose an experimental investigation of finding stable black holes in underground laboratories. 
 (iii) Relativistic black holes produced by cosmic rays \cite{cosmic} passing through stars or planets may 
evaporate to minimal size or increase in size by accretion depending on the interaction cross section. 
We estimate what is the lowest energy of incoming cosmic rays that allows black holes to increase in 
size and accrete all available matter in a star. 

\section{Lovelock black holes in odd dimensions}

As we mentioned in the introduction our investigations are largely independent of the choice of a specific 
gravity model. We only assume that black holes have a minimal mass, at which the Hawking temperature 
vanishes.  Lovelock black holes \cite{lovelock} in odd dimensions are prime examples satisfying these 
constraints.   Gravitational theories including Lovelock terms are perfectly acceptable, as their field 
equations contain only second order derivatives and satisfy standard requirements on a classical theory 
of gravity just as much as Einstein gravity does.  The langrangian of the $n$th order Lovelock term, $L_n$, 
is built from an antisymmetrized combination of the $n$th power of the curvature tensor.  In $D$ dimensions 
a linear combination of Lovelock terms up to $n=(D-1)/2$, such as
\begin{equation}
L=R+\sum_{n=2}^{(D-1)/2} M_P^{2(n-4)}\alpha_n L_n 
\label{lagrangian}
\end{equation}
contribute non-trivially to the equations of motion. Note that $n=0$ corresponds to the cosmological 
constant, the $n=1$ term is Einstein gravity, and Gauss-Bonnet gravity corresponds to $n=2$.  We extract 
an appropriate power of the Planck mass to make the coupling constants $\alpha_n$ dimensionless. 
Also note that we use $M_P$ as a unknown quantity, dependent on $D$, to be determined by experiments.  
We also keep the coupling constants, $\alpha_n$ as free parameters, but when we need to make an 
estimate of an observable quantity we will often consider predictions under the assumption 
$\alpha_n=O(1)$, or, alternatively $M_{\rm min}\simeq M_P$, where $M_{\rm min}$ is the minimal mass.  
We use a definition of Newton's constant in terms of the Planck mass of the "Particle Data Book" 
\cite{databook},
\begin{equation}
8\pi G_D=(2\pi)^{D-4}M_P^{2-D}  
\end{equation}
With this definition the experimental lower bound on $M_P$ is $M_P\geq 1$ TeV \cite{databook}. 
 This bound was calculated assuming Schwarzschild-Tangherlini black holes. We expect that the 
lower bound on the Planck mass when black holes have a minimal mass is somewhat lower. 
The reason for that is the small amount of radiation produced when the black holes approach the 
minimal mass.   Since it barely radiates one could only recognize its production by the missing energy 
that could have been taken away by other neutral particles. 

Analytic spherically symmetric black hole solutions have been found  for $n=2$ 
\cite{boulware}\cite{wiltshire}\cite{myers} and 3 \cite{dehghani} Lovelock gravity. While in even 
dimensions black hole masses have no lower bound, in odd dimensions the mass is limited by the 
Lovelock coupling $\alpha_n$, to ensure the existence of a horizon.  In particular for $D=5$ and 7
\begin{equation}
(\mu^{(D)}M_P)^{D-3}\geq(D-3)!\alpha_{(D-1)/2}.
\label{bounds}
\end{equation}
 In (\ref{bounds}) $\mu^{(D)}$ is the radius of horizon for the $D$-dimensional Tangherlini black hole 
\cite{giddings}
\begin{equation}
\mu^{(D)}=\frac{1}{M_P}\left(\frac{k_DM_{\rm ADM}}{M_P}\right)^{1/(D-3)},
\label{mass}
\end{equation}
where
\begin{equation}
k_D=\frac{2(2\pi)^{D-4}}{(D-2)\Omega^{(D-2)}},
\end{equation}
and
\begin{equation}
\Omega^{(D-2)}=\frac{2\pi^{(D-1)/2}}{\Gamma[(D-1)/2]}.
\end{equation} Using the latter equations the lower bounds on the ADM mass are 
\begin{eqnarray}
M&> &M_{\rm min}^{(5)}=3\pi\alpha_2M_P,\nonumber\\
M&> &M_{\rm min}^{(7)}=\frac{15}{2}\alpha_3M_P
\end{eqnarray}
for $D=5$ and $D=7$, respectively. It is worth noting that the minimal mass is larger than the Planck mass 
if the coupling of the Lovelock terms, $\alpha_n=O(1)$. This result may be of significance if we consider 
the stability of minimum masses under quantum gravity. It is possible that quantum effects destabilize 
classically stable black holes. In what follows, we assume that even if that were true, their lifetimes are 
long enough to leave our classical calculations valid. 

The radius of horizon is 
\begin{eqnarray}
r_h^{(5)}&=&M_P^{-1}\sqrt{(\mu^{(5)}M_P)^2-2\alpha_2},\nonumber\\
r_h^{(7)}&=&M_P^{-1}\sqrt{-6\alpha_2+\sqrt{(\mu^{(7)}M_P)^4-24\alpha_3+36\alpha_2^2}},
\label{radius}
\end{eqnarray}
in 5 and 7 dimensions, respectively.

The Hawking temperature in 5 dimensions is
\begin{equation}
T_H=M_P\frac{\sqrt{(\mu^{(5)}M_P)^2-2\alpha_2}}{2\pi ((\mu^{(5)}M_P)^2+2\alpha_2)}.
\label{hawking}
\end{equation}
In $D$=7 the expression for $T_H$ is much more complicated, but also proportional to the radius of 
horizon, as in $D=5$. The reason for this is simple. In odd dimensions the metric functions are even 
functions of the radius and the surface gravity is proportional to $dg_{tt}/dr$. Then both of these 
theories the radius of horizon and the Hawing temperature vanish as $\sim\sqrt{M-M_{\rm min}}$. 
This is a general feature of theories we consider in this paper. 

 Strictly speaking minimum mass black holes do not exist.  Scalars, like the curvature scalar, square 
of the Ricci tensor, and the Kretschmann scalar all diverge at zero radial coordinate, when the 
radius of the horizon vanishes. This state is, however, never reached in the semi-classical theory 
because, as is shown by (\ref{timedependence}), it takes infinite time for the mass of a black 
hole produced with $M>M_{\rm min}$ to reach  $M_{\rm min}$ . 

While at sufficiently large mass, $M>M_P$, the Hawking temperature approaches its value in Einstein 
gravity $T_H\simeq(2\pi r_h)^{-1}$, at $M\simeq M_{\rm min}$ in Lovelock gravity it can be 
parametrized by the constants, $M_P$ and $\alpha_n$ as
\begin{eqnarray}
T_H^{(5)}&\simeq &\frac{M_P}{\sqrt{2\alpha_2}}\sqrt{\frac{M}{M_{\rm min}^{(5)}}-1}=\sqrt{\frac{3\pi}{2}}\frac{M_P^{3/2}\sqrt{M-M_{\rm min}^{(5)}}}{M_{\rm min}^{(5)}}\nonumber\\\
T_H^{(7)}&\simeq &\frac{M_P}{3}\sqrt{\frac{2\alpha_3}{\alpha_2}}\sqrt{\frac{M}{M_{\rm min}^{(7)}}-1}
\end{eqnarray}
In general we may parametrize the dependence of the Hawking temperature on the minimal mass 
by the two parameters $M_s$ and $M_P$ as
\begin{equation}
T_H\simeq \,M_s  \sqrt{\frac{M}{M_{\rm min}}-1},
\label{hawking2}
\end{equation}
 where $M_s ,M_{\rm min}=O(M_P)$.  We will use this  formula in all subsequent calculations. However, 
whenever we need concrete numbers for calculating a particular phenomenon related to stable black 
holes we will rely on the 5 dimensional formulas, which, hopefully, give a good order of magnitude 
estimate in higher dimensional spaces, as well.
 
 \section{Stable black holes as components of dark matter}
If black holes become stable after approaching a nonzero minimal mass then their number can only 
be reduced by collisions with each other, when the mass of one of the black holes is converted into 
radiation. Therefore, most primordial black holes and those produced later by cosmic rays should still 
be around at the present time.   For large black holes the higher dimensional Lovelock terms become 
irrelevant and the Schwarzschild approximation is valid.  Thus, the spectrum of black holes produced 
after inflation is unaffected by them.  Hawking and Page \cite{hawking} have shown that primordial 
black holes lighter than $10^{15}$ g would have decayed by now.  However, in odd dimensional 
Lovelock theory they would have all ended up in the near-minimal mass state.   These black holes, 
along with small black holes produced later, would form part of dark matter.  In the current section 
we investigate the effect of such a component of dark matter and derive bounds on their minimal mass, 
$M_{\rm min}$, and their relative abundance compared to other components of dark matter. 
 
First we show that for a very wide range of the Planck mass, $M_P$, and of minimal mass, $M_{\rm min}$, 
the passage of non-relativistic black holes through matter follows a simple pattern: A minimal mass black 
hole accretes a parton or an electron increasing the mass and the Hawking temperature of the black 
hole. The accretion is followed by the rapid decay of the black hole towards the minimal mass state, 
very probably before the next accretion takes place.  This sequence of events allows us to derive a 
recursion relation for the velocity distribution and an integral equation for the limiting distribution. Not 
surprisingly, in view of the central limit theorem, the solution of the integral equation is a Maxwell distribution. 
It is interesting that the average velocity in the limiting distribution, depending on $M_{\rm min},$ 
can be higher than average velocity of incoming dark matter particles.  Using the velocity distribution we 
show that in neutron stars the near-stable black holes are confined to a sphere of radius $r\lesssim 100$ m 
 where they reach an equilibrium density in a few decades after the formation of the neutron star.   In 
equilibrium, the number of black holes annihilated in pairwise collisions is equal to the number of black 
holes captured by the neutron star from dark matter.  We will also show that black holes cannot be 
captured by ordinary stars, if $M_{\rm min}=O$(1 TeV), as  the velocity of most of black holes exceeds 
the escape velocity.  However, if $M_{\rm min},\,M_P\gtrsim 100$ TeV, then their velocity is smaller and 
 the captured black holes accumulate inside the sun, as well.
 
At the end of this section we will investigate the experimental signature of stable black holes passing 
through experimental devices. The signature is quite universal. It is in large measure independent of the 
minimal mass, up to $M_{\rm min}\simeq10^5$ TeV, and of the velocity of the black holes.
 
\subsection{Decay of near minimal mass black holes in vacuum}
When we calculate the decay of black holes we can use the Schwarzschild metric for black holes with 
$M>>M_{\rm min}$,  while  (\ref{hawking2})  can be applied to black holes with $M\lesssim 2M_{\rm min}$. 
Considering that standard model particles live on a 3 brane the radiation happens mainly on the brane 
\cite{emparan}. Then the decay rate of near minimal  black holes in vacuum is 
\begin{equation}
\frac{dM}{dt}=-g\frac{ \pi^2 }{120}4\pi r_h^2T_H^4=-g\frac{\pi^3}{30}M_s^2\left(\frac{M}{M_{\rm min}^{(5)}}-1\right)^3
\label{decayrate}
\end{equation}
where $g$ is the number of effective degrees of freedom. In 7 dimensions an extra factor of 2/9 should 
be included on the right hand side of (\ref{decayrate}).  Note that, unlike for Schwarzschild black holes, in 
a theory with stable black holes the contribution of emission into the bulk is negligible when $M$ is close to 
$M_{min}$. The contribution of bulk radiation to (\ref{decayrate}) would have higher powers of $r_h$ and 
$T_H$, both vanishing when $r_h\to 0$.

Integrating (\ref{decayrate}) gives
\begin{equation}
M(t)=M_{\rm min}+\frac{\Delta M}{\sqrt{1+t\frac{\pi^3 g M_s^2}{15  M_{\rm min}^3}(\Delta M)^2  }},
\label{timedependence}
\end{equation}
where $\Delta M= M(0)-M_{\rm min}$ is the energy gained by the absorption of a parton.  The half-time 
(time required to reach $M(t_{1/2})=(M(0)+M_{\rm min})/2$ is
\begin{equation}
t_{1/2}\simeq 0.06\frac{M_{\rm min}}{M_s^2}\left(\frac{M_{\rm min}}{\Delta M}\right)^2.
\label{halftime}
\end{equation} 
When non-relativistic black holes travel through matter they absorb partons from atomic nuclei. Since they 
cannot leave colored matter behind, they must pull out an antiparton from the chromosphere of the black 
hole, as well.  Altogether, they only gain about a mass of $\Delta M=O(0.5$GeV). Note that $\Delta M$ is, 
in a very good approximation, independent of $M_{\rm min}$. When, following the accretion process, 
owing to the limited available total mass, $\Delta M$, they can only emit light degrees of freedom, mostly 
photons, electrons, positrons, neutrinos and antineutrinos, with an occasional muon or pion. Then the 
effective number of degrees of freedom, including greybody factors \cite{anchordoqui} is $g\simeq 23\,
\, 2/3$.  Note that for 7 dimensional black holes the prefactor in (\ref{halftime}) should be decreased to 
0.03. 

Applied to 5 dimensional black holes we obtain the following expression for $t_{1/2}$
\begin{equation}
t_{1/2}=0.013\frac{M_{\rm min}^2}{M_P^3}\left(\frac{M_{\rm min}}{\Delta M}\right)^2\simeq 5.2\times10^4\,{\rm TeV}^{-1}\frac{M_{\rm min}^4}{M_P^3}
\label{fivelifetime}
\end{equation}
Note that here and all in subsequent equations $M_P$ and $M_{\rm min}$ are measured in units of TeV 
and are therefore dimensionless.

As we will see below, the decay time (\ref{fivelifetime}) is much shorter than the average time between collisions.  
Then it follows that the passage through ordinary matter is a series of accretions of partons each followed by 
almost instantaneous decay back to near minimal mass. 

\subsection{Cross section of TeV scale black holes}
For the sake of definiteness we will consider black holes in a RS I scenario.  A classical calculation of the 
Bondi-Hoyle-Littleton accretion can be performed by considering the geodesics of pointlike incoming 
particles at impact parameter $b$.  If the impact parameter is smaller than the compactification radius 
then the absorbtion cross section is defined by the maximal $\pi b_{\rm max}^2$, where $b_{\rm max}$ 
is the maximal impact parameter at which the projectile is captured by the black hole.  For this calculation 
we must use the $D$ dimensional Einstein-Gauss-Bonnet metric  in the neighborhood of the black hole.  
We obtain for $D=5$
\begin{equation}
\sigma= 4\alpha_2 \pi v^{-2} M_P^{-2}\left(\frac{1}{2}+\frac{v}{\gamma}\right)=\frac{4}{3\, v^2}M_{\rm min}M_P^{-3}\left(\frac{1}{2}+\frac{v}{\gamma}\right),
\label{cross}
\end{equation}
where $v$ is the relative velocity and $\gamma$ is the corresponding Lorentz factor. Since we are 
interested in order of magnitude estimates we use classical cross sections throughout this paper.

The increase of $\sigma$ with decreasing $v$ is due to the increase of the maximal impact parameter 
at which the particle is captured.  However, at large impact parameters the 5 dimensional metric is no 
longer applicable, because at large distances from the black hole the metric becomes effectively 4 
dimensional, with standard 4 dimensional gravitation strength, which hardly causes any deflection of 
particles. The transition between the 5 and 4 dimensional behaviors happens at $r\simeq m_r^{-1}$ 
where  $m_r$ is the radion mass.  The largest lower bound on the radion mass, 120 GeV,  has been 
derived in \cite{lower}. We are not aware of upper bounds.

To incorporate the low velocity cutoff imposed by the radion mass we modified our calculation of the 
cross section by using the simple minded assumption that the metric reduces to Minkowski metric at 
distances larger than $r_{\rm max}=m_r^{-1}$. Then at $m_r<<M_P$ and $v<<1$
\begin{equation}
\sigma\simeq\frac{\pi}{m_r^2}.
\label{sigrad}
\end{equation}

For larger black holes the cross section can be calculated from the Schwarzschild metric. One obtains 
\cite{unruh}
\begin{equation}
\sigma=4\pi \mu_D^2,
\end{equation}
where $\mu_D$ is given in (\ref{mass}). In particular, for $D=5$
\begin{equation}
\sigma=\frac{4}{3} M M_P^{-3}
\label{largemass}
\end{equation}

\subsection{Mean free path of black holes}

The mean free path in a medium is given by $l=(\sigma \rho)^{-1}$, where $\rho$ is the number density 
and $\sigma$ is the collision cross section.  To get a rough estimate of the mean free path we will use 
the absorption cross section. We will  discuss neutron stars first.  

Nuclear matter has the parton number density, with about 3 partons per neutron
\begin{equation}
\rho=4\times 10^{-12}{\rm TeV}^3.
\end{equation}

Then using (\ref{cross}) and (\ref{sigrad}), we obtain
\begin{equation}
1.1 \times10^9 \,{\rm TeV}^{-1}\leq\ell_n =\frac{1}{\rho \sigma}\leq3.8\times 10^{11}\,  {\rm TeV}^{-1}\frac{M_P^3}{M_{\rm min}}
\label{nuclear}
\end{equation}
where the upper limit is given by the minimum of (\ref{cross}), at $\gamma\to\infty$ and the lower limit 
by (\ref{sigrad}) at the minimum radion mass \cite{lower} $m_r=$120 GeV. However, if $M_P > 10$ TeV, 
then the velocity dependent classical formula gives a larger lower bound then the one obtained using the 
minimum radion mass ($\langle v^2\rangle=x \,\Delta M \,M_{\rm min}^{-1}$ will be calculated in the 
next subsection): 
\begin{equation}
3.8\times 10^{11}\,  {\rm TeV}^{-1}\langle v^2\rangle\frac{M_P^3}{M_{\rm min}}=2\times 10^{8}\,  {\rm TeV}^{-1}x\frac{M_P^3}{M_{\rm min}^2}\leq\ell_n
\label{nuclear2}
\end{equation}
The lower bound (\ref{nuclear2}) ensures that black holes undergo many collisions in nuclear matter (or in stars), 
except in the case when our mass parameters, $M_P$ and $M_{\rm min}$ are at the upper end of the range of 
interest, near $M_P\sim M_{\rm min}\gtrsim 10^{15}$.  Then the mean free path becomes larger than a few km 
and most black holes undergo a small number of collisions only.  Since the number of collisions inside the star 
largely depend on the total mass only,  a similar statement applies to a star, like our sun, as well.

The corresponding bounds on the mean free path in water are
\begin{equation}
 600\,{\rm km}\frac{M_P^3}{M_{\rm min}}\geq \ell_w \geq
\begin{cases}
2.9 \,{\rm km}\\
0.3\,{\rm km}\,\,x\,\frac{M_P^3}{M_{\rm min}^2}
\end{cases}
\label{water}
\end{equation}
where the first lower bound is obtained using cross section (\ref{sigrad}), while the second using (\ref{cross}) 
at low $v$. These numbers will be important when we will discuss experimental constraints on $M_{\rm min}$ 
and $\delta$, the fraction of dark matter in the form of microscopic stable black holes.

While passing through earth black holes undergo a large number of collisions, unless $M_P>$ is very large.  
The limits for water, (\ref{water}), should be divided by 5.5. Even though they collide with nuclei many times 
inside the earth, the earth cannot capture black holes, because, as we show in the next subsection, their 
average velocity, $\langle v\rangle=O(10^{-2}\,c)$ is orders of magnitude higher than the escape velocity. 

Let us compare now the decay time of black holes with the collision time in nuclear matter.   The average 
time between collisions is $t_c= \ell_n /v$.  In the next subsection we will determine the average speed of 
black holes, dependent only on having a large number of collisions and largely independent of the  matter 
density.  We will obtain $v\simeq 0.5 \sqrt{\Delta M\, \rm /M_{\rm min}}\simeq0.01\,M_{\rm min}^{-1/2}$. 
Using (\ref{nuclear}) and   taking the accreted energy $\Delta M\simeq0.5$ GeV in each collision, we 
obtain the following expression for the ratio of collision time and decay time in nuclear matter
\begin{equation}
6.8\times 10^9 \frac{M_P^6}{M_{\rm min}^{9/2}} \geq\frac{t_c}{t_{1/2}}\geq
\begin{cases}1.8\times 10^8\frac{M_P^3}{M_{\rm min}^{7/2}} \\
3.8\times 10^7\,x\, \frac{M_P^6}{M_{\rm min}^{11/2}}
\end{cases}
\label{ratio}
\end{equation}
Taking the  root of the product of the two lower limit, setting $M_P\simeq M_{\rm min}$, and using  the very  
conservative value $x=0.2$ we obtain a lower limit that is independent of $M_P$,  namely 
$t_c\, / t_{1/2}\geq9\times 10^{7}.$ Even in the center of the neutron star where the density is 2-3 times 
higher than that of nuclear matter the ratio is much larger than 1.  

The important message in (\ref{ratio}) is that the passage of black holes through any medium, including 
the core of neutron stars, follows a simple pattern. After a $M\simeq M_{\rm min}$ black hole collides  
with  and absorbs a parton its excess mass evaporates long before the next collision to become a 
$M\simeq M_{\rm min}$  black hole again.  This sequence will be important in the next subsection in 
which we investigate the velocity  distribution  of non-relativistic black holes close to their minimum mass 
in matter.

\subsection{Diffusion of black holes in matter}

We consider stable black holes arriving, as part of the dark matter, at a star or a planet, in which  they 
undergo multiple collisions. For simplicity, we will consider the star as infinite matter. If the temperature of 
the star is low compared to the kinetic energy of incoming stable black holes one may naively think that 
the black holes lose velocity, until their kinetic energy reaches the ambient temperature.  In fact, this is not 
the case at all because the inelastic collision and subsequent decay of the black holes may result in, 
depending on the mass of the black hole, an increase of the average velocity.  The black holes are not 
in equilibrium with the surrounding matter, because of the long collision time compared to the decay 
time, as shown by (\ref{ratio}). When the black hole accretes a parton then, as an average, it slows down 
slightly in the star rest system. However,  after accretion it decays soon approaching its minimal mass state 
by emitting a small number of particles and gaining extra velocity from the recoil.  After multiple collisions 
the addition of these recoil velocities results in a random velocity distribution, which is independent of the 
original velocity distribution of incoming dark matter particles. The average  velocity depends on 
$M_{\rm min}$ but as we will see below if $M_{\rm min }\lesssim 1000$ TeV then it is higher than the 
average velocity of dark matter particles. 

Let us consider the first step of the repeating two-step process, the accretion. As black holes accrete 
partons of energy $\Delta M$ then, as an average, their momentum is conserved in the star frame but 
they lose approximately the fraction $\Delta M/M_{\rm min}$  of their velocity. However, in the second 
step, when they subsequently radiate essentially all of the accreted $\Delta M$ energy they also gain 
momentum 
\begin{equation}
 p_+=- \sum_i ^nq_i,
 \end{equation}
 where $q_i$ are the momenta of emitted particles. The $n$ particles are emitted isotropically. Since the 
Hawking temperature  $T_H>>\Delta M$ the individual particle momentum distribution can be 
approximated by phase space.  Neglecting the mass of emitted particles the momentum gained by the 
black hole after the end of Hawking radiation is
 \begin{equation}
f^{(n)}( p)=N^{-1} \int \delta\left(\vec p+\sum_i^n \vec q_i\right)\delta\left(\Delta M-\sum_i^n q_i\right)\prod_i ^nd^3q_i,
 \label{momentum}
 \end{equation}
 where $q_i=|\vec q_i|$ and $N$ is a normalization constant.  Note that $n$, the number of particles 
emitted, is small, $\langle n\rangle\sim 2$, as it can be ascertained using the infinite temperature limit 
of the Bose-Einstein and Fermi-Dirac distributions.  In particular,  
 \begin{eqnarray}
 f^{(1)}(p)&=& \frac{1}{4\pi \Delta M^2}\delta(\Delta M-p),\nonumber\\
  f^{(2)}(p)&=& \frac{1}{N}\left[\log\left(\frac{\Delta M+p}{\Delta M-p}\right)-\frac{p}{12\Delta M}\left(21+(p/\Delta M)^2\right)\right]
  \end{eqnarray}
  where $N=\Delta M^3\pi [\log(4096)-17]/36$.  The distribution $f^{(3)}(p)$ is slightly more complicated, 
containing polylogarithms. 
  
Now the question is whether after successive accretions and subsequent Hawking radiations the momentum 
distribution approaches a limiting distribution or not. Originally the speed of the black hole is about 
$v\simeq 10^{- 3}c$, a typical speed for dark matter particle. The change of the speed of the black hole 
due to the emission of particles is $\delta v= O(p/M_{\rm min})=O(\Delta M/M_{\rm min})$. This is of the 
same order of magnitude as the original velocity, at least if the mass of the black hole is $M=O($1TeV).   
Even if the black hole has mass $M_{\rm min}>>1$TeV it loses memory of the original direction,  as it 
undergoes a very large number of collisions. Thus, its distribution becomes isotropic (discounting the effect 
of the gravitational force of the star). After a few collisions the speed distribution of the black hole in 
homogeneous infinite matter approaches an asymptotic form determined by the integral equation
  \begin{equation}
  g(v)=\frac{1}{N} \int g\left(\vec v[1+\Delta M/M_{\rm min}]-\vec q/M_{\rm min}\right)f(q)d^3q
  \label{recursion}
  \end{equation}
  where the multiplier $1+m/M_{\rm min}$ is the factor by which an accretion  slows down the black hole 
and $f(q)$ is the normalized weighted average of distributions $f^{(k)}(q)$ over $k$.
  
As expected from the central limit theorem, the solution of (\ref{recursion}), in leading order of  
$\Delta M/M_{\rm min}$, is a Gaussian distribution,
  \begin{equation}
  g(v)=\left(\frac{3 \Delta M\, M_{\rm min}}{2\pi\langle q^2\rangle}\right)^{3/2} \exp\left(-\frac{3\Delta M\,M_{\rm min}\,v^2}{2\langle q^2\rangle}\right),
  \label{distribution}
  \end{equation}
  where
  \begin{equation}
  \langle q^2\rangle=\int q^2 f(q)d^3q = x\,\Delta M^2,
  \end{equation}
  with $x\leq1$. For massless particles the value of $x_n$ ($x$ for $n$ decay products) is $x_n=(1+3(n-1)/4)^{-1}$. 
  
  The expectation value of the squared velocity is  \begin{equation}
  \langle v^2\rangle= x \frac{\Delta M}{M_{\rm min}}\simeq 5\times 10^{-4}x\, M_{\rm min}^{-1}
  \label{averagevelocity}
  \end{equation}
One would expect that the number emitted particles is small, so $1>x\gtrsim 0.2$, leading, for 
$M_{\rm min}\simeq$ 1 TeV, to an average velocity, $v_{\rm av}\gtrsim 10^{-2}c$. In other words, the 
multiple collision process  {\em accelerates}  the minimal mass black hole. provided $M_{\rm min}\lesssim 1000$ TeV.
  
 Eq.  (\ref{distribution}) is a Maxwell distribution with $kT\lesssim 200$ MeV.  It is interesting to observe that 
the temperature of the gas of black holes is independent of the temperature of the medium. The two different 
temperatures are maintained by the peculiar nature of the interaction of black holes with particles in the medium. 
  
  \subsection{ Near minimum mass black holes in stars}
  
  The calculation of the average velocity in the previous section has assumed that no external forces act on 
the black hole. This is not true in stars and particularly in a neutron star in which  the gravitational field is 
enormous. Using a radius of $R=$10 km, and  mass of $M_n=1.5 \times M_{\rm sun}$ we obtain the 
gravitational potntial, $g$, as 
  \begin{equation}
  g=\frac{G M_n }{R}=0.2 >> \frac{v^2}{2}\lesssim 2.5\times10^{-4}. 
  \label{potential}
  \end{equation}
  
The velocity of black holes is then much smaller than the escape velocity and the black holes start to move 
towards the center of the neutron star.  At the center, however, the gravitational field vanishes, so the black 
holes rapidly reach the velocity distribution described in the previous subsection.  They will be confined to a 
sphere the approximate radius of which is obtained from equating the average random velocity with the 
escape velocity on the surface of this sphere. To determine this radius we need a model for the density 
distribution.  We choose a normalized quadratic density distribution of
  \begin{equation}
  \rho(r)=M_n\frac{30}{4\pi R^3}\left(1-\frac{r}{R}\right)^2.
  \end{equation}
  Other choices would slightly change the final result.
  Then near the center the gravitational potential is
  \begin{equation}
  g(r)\simeq  M_n  G \frac{10 \,r^2}{R^3}.
  \end{equation} 
  Then using  (\ref{averagevelocity}) and (\ref{potential}) we obtain the the radius, $r_c$, of the 
aforementioned sphere  from
  \begin{equation}
  \frac{\langle v^2\rangle}{2}=g(r)=2 \frac{r^2}{R^2}
  \end{equation}
  We obtain 
  \begin{equation}
  r_c=110\, {\rm m}\,\sqrt{x}\,M_{\rm min}^{-1/2}.
  \label{centralradius}
  \end{equation}
  
  Inside the sphere of radius $r_c$ black holes annihilate by colliding with each other.  In an equilibrium 
state the number of annihilated black holes must equal the number of black holes captured by the 
neutron star from dark matter and from those produced by cosmic rays at the surface of the star.
  
  If minimum mass black holes constitute a fraction $\delta$ of dark matter, then using the dark matter 
density of 300 GeV $m^{-3}$, a radius of 10 km , and a mass of 1.5 times that of the sun we obtain the 
following capture rate of black holes traveling at $\sim 270$ km/s
  \begin{equation}
  N_{\rm cap}=2.5 \times 10^{16}s^{-1}\, \delta M_{\rm min}^{-1}
  \label{capturerate}
  \end{equation}
  
  The number of black holes undergoing two-body collisions  in the central sphere per unit time and unit 
volume is
  \begin{equation}
  N_a=\rho_{\rm BH}^2 \sigma v
  \end{equation}
  Then using the Maxwell distribution (\ref{distribution}) the average relative velocity of two black holes is
  \begin{equation}
  v=4 \sqrt{\frac{x}{3\pi}}\left(\frac{\Delta M}{M_{\rm min}}\right)^{1/2}
  \end{equation}
  Now after attaining equilibrium $V N_a=N_{\rm cap}$ so the density of black holes in the central sphere is
  \begin{equation}
 \rho_{\rm BH}=\frac{N_{\rm cap}}{\sqrt{V\sigma v}}=18\, {\rm m}^{-3}\,\frac{\delta^{1/2}}{x \sqrt{\sigma /{\rm m}^2}}M_{\rm min}^{1/2}
  \end{equation}
  where $V$ is the volume of the central sphere confining the black holes. The cross section is measured in m$^2$.
  
  Now we can use our three bounds on $\sigma$ to find bounds on the density of black holes in the central 
region of a neutron star.  Using (\ref{cross}) in the extreme relativistic limit we obtain an upper limit, while 
using (\ref{cross}) with $\langle v^2\rangle$ taken from (\ref{averagevelocity}) provides one of the lower 
bounds, and finally (\ref{sigrad}) provides another lower bound, as follows
  \begin{equation}
1.8\times 10^{19} \,{\rm m}^{-3}\frac{\sqrt{\delta}}{x}\,M_P^{3/2}\geq \rho_{\rm BH}\geq\begin{cases}
 4\times 10^{17}\,{\rm m}^{-3}\sqrt{\frac{\delta}{x}}\,M_P^{3/2}\,M_{\rm min}^{-1/2},\\
  2.9\times 10^{18}\,{\rm m}^{-3}M_{\rm min}^{1/2}.
  \end{cases}
  \label{lower1}
  \end{equation}
These limits can be compared with the density of partons in the central region, $\rho_p\simeq 10^{45}m^{-3}$.
   
   How long a time is needed to fill a newly created neutron star with black holes at equilibrium density? The 
total number of black holes in the central region divided by the rate of black hole capture provides two 
lower bounds on $t_{\rm fill}$, obtained from (\ref{lower1}) 
 \begin{equation}
 t_{\rm fill}\geq V \frac{\rho_{\rm BH}}{N_{\rm cap}}\geq 
\begin{cases} 
20\left(\frac{x}{\delta}\right)^{1/2}\,{\rm years}\\
3 \frac{x}{\sqrt{\delta}}\,M_P^{3/2}\,M_{\rm min}^{-1}\,{\rm years}
\end{cases} 
 \end{equation}
 It is worth noting that if $M_P$ and $M_{\rm min}$ are close to $M_P^{(4)}\simeq 2.4\times 10^{15}$ TeV 
then the process of "filling" the neutron star with black holes, which, according to (\ref{centralradius}), 
are now concentrated at the very center of the neutron star, takes tens of millions of years.  This would 
imply a change in the surface temperature on the scale of tens of millions of years, as well. An 
investigation of time dependence of the surface temperature and its possible observation in neutron stars 
has been performed assuming that WIMPs constitute dark matter \cite{kouvaris}.  Black holes as 
components of dark matter, could effect the surface temperature of neutron stars considerably less than 
WIMPs.  The mass of the black hole produced from the collision of two black holes may be substantially 
less than the sum of the masses of the colliding black holes \cite{bekenstein}.  The lowest possible 
resultant mass is obtained from adding entropies, rather than masses.  In the case of near minimal mass 
stable black holes that would imply a third black hole of near minimal mass, as well.  The rest of the 
energy, $\Delta E\simeq M_{\rm min}$, would be dissipated in the form of gravitational radiation, which 
would escape the neutron star. In such an extreme case the annihilation of stable black holes would not 
lead to the heating of stars, at all.   However, if the collisions are not entropy conserving then the masses 
of the black holes may simply add. In the latter case a large fraction of the mass would be radiated 
in the form of elementary particles.  We intend to return to this question in a future publication. 
   
    We will consider ordinary stars briefly.  In a star like our sun, or in any star, for that matter, black holes 
undergo a large number of collisions to reach the equilibrium velocity distribution (\ref{distribution}).  
Owing to the large size of the sun or other ordinary stars, the surface gravity is smaller then the average 
squared velocity, provided the minimum mass is smaller than $M_{\rm min}< 100$ TeV,
   \begin{equation}
   g=\frac{M_{\rm sun}G}{R_{\rm sun}}=1.9\times 10^{11} ({\rm m/s})^2<< \frac{1}{2}\langle v^2\rangle\simeq 1.8 \times 10^{13} x \,M_{\rm min}^{-1}({\rm m/s})^2.
   \end{equation}
   However, if $M_{\rm min}>100$ then black holes get captured in the sun, as well. In fact, the number 
of black holes collected in a central sphere is much larger than those in a neutron star, owing to the sun's 
much larger cross section to capture dark matter.  We intend to address this question along with the 
question of dissipation  of heat produced from the annihilation of black holes from stars \cite{kouvaris} 
   in a future publication. 

   \subsection{Signature of stable black holes traversing matter and bounds on the minimal mass and abundance}
 
 As we pointed out in the previous subsection, stable black holes contributing to dark matter travel through 
visible matter in an unusual manner. Suppose at a given instant the black hole is close to its minimal mass with 
nearly zero Hawking temperature.  Then colliding with an atomic nucleus, or an electron it can absorb a 
parton or an electron and increase its mass and Hawking temperature. Note that stable black holes near the 
minimum mass have positive specific heats. The mass of the black hole increases by the energy of the 
accreted object, $\Delta M<< M_{\rm min}$,  while, as shown by (\ref{hawking2}), its temperature increases 
to $T_H\sim \sqrt{M_{\rm min} \Delta M}$. This implies that the increase in temperature is larger by at least 
two orders of magnitude than the increase of energy.  The black hole becomes a very hot object with very 
low heat capacity. 
 
 It is difficult to calculate the average particle number and momentum distribution in this strange system.  
However, owing to the fact that the total energy to be radiated isotropically is in the range of 
$\Delta M\simeq0.1-0.5$ GeV the decay products are restricted to light particles, mostly photons and 
electrons and neutrinos, with a possible muon or pion.  A certain amount of gravitons are also emitted, 
further decreasing the missing energy \cite{graviton}.  Due to phase space considerations lower multiplicity 
is probably preferred, so most of the events should contain 1-3 visible particles.  As the recoiling black hole 
takes up momentum, the total momentum is not balanced even if we disregard the momentum taken 
away by gravitons and neutrinos.  In many ways the events should be similar to the decay of a proton-like 
object with a mass of less than a GeV. Therefore, the ideal experiment searching for stable black holes is 
in large underground detectors.  
 
 As in proton decay experiments, the most important background signal is the inelastic interactions of 
atmospheric neutrinos. In fact, the frequency of $E< 1 GeV$ $\nu_e,\, \bar\nu_e$ events serves a rough 
upper bound on the frequency of stable black hole collisions.  In the Minnesota proton decay 
experiment \cite{minnesota}  69 neutrino events, resembling proton decay were found in 3300 m$^3$ 
fiducial volume in 80 days. As stable black hole emission events also resemble proton decay events, 
we may use the frequency of those neutrino events as an upper bound on black hole evaporation events. 
This amounts to an upper bound on the event rate of  $N_e=3\times 10^{-9}$ m$^{-3}$s$^{-1}$. Compare 
this with the limits on the rate of predicted stable black hole events, using our lower and two upper 
bounds on the cross section :
 \begin{equation}
162 \,\delta    \,M_{P}^{-3}\,{\rm m}^{-3}{\rm s}^{-1}  \leq N_{\rm BH}\leq
\begin{cases}
6200\,\delta  \,M_{P}^{-1}{\rm m}^{-3}{\rm s}^{-1},\\
 3.7\times 10^5 \frac{\delta}{x} M_{\rm min} \,M_{P}^{-3} {\rm m}^{-3}{\rm s}^{-1}
 \end{cases}
 \label{eventrate}
\end{equation}
Then the lower bound on the black hole decay rate provides a lower bound on the Planck mass that 
must be satisfied so that we would not contradict the observed rate of neutrino events resesmbling proton decay
\begin{equation}
\delta \,M_{P}^{-3}<2 .\times10^{-11}.
\label{lower3}
\end{equation}

The upper bounds on the black hole decay rate provide two further conditions. If  either of these are 
satisfied then constraint on the event rate is certainly satisfied. The two conditions are
\begin{equation}
\delta \,M_{\rm min}^{-1}\leq 
 5 \times 10^{-13}
  \label{upper1}
\end{equation}
and
\begin{equation}
\frac{\delta}{x} \frac{M_{\rm min}}{M_P^3}\leq 
 8 \times 10^{-15}
 \label{upper2}
\end{equation}

Then (\ref{lower3}) implies that if the {\em dark matter consists solely of black holes}  then $M_P\geq 3700$ TeV. 
The bounds (\ref{upper1}) and (\ref{upper2}) imply that the event rate is certainly low enough if 
$M_{\rm min}\geq 2\times 10^{12}$ TeV or  $M_P^{3/2}M_{\rm min}^{-1/2}\geq 1.1\times\, \sqrt{x}\,10^7$ TeV.  
If $M_P\simeq M_{\rm min}$ then the lower bound on  $M_{\rm min}$ is somewhere in the range
\begin{equation}
3.7 \times 10^3 {\rm TeV }\leq M_{\rm min}\leq 1.1\times 10^7 {\rm TeV}.
\end{equation}
Black holes with a minimal mass below the lower bound would surely have been detected. If the minimal 
mass is above the upper bound then the black hole would certainly have not been detected.

{\em Unless only a small fraction of dark matter is made up from stable black holes the minimum mass of 
black holes must be much larger than 1 TeV.} If the mass of black holes is very large, 
$M_{\rm min}>10^{10}$ TeV, then the decay event is not any more localized.  Using 
(\ref{timedependence}) in 5 dimensions we obtain for the expected remaining mass as a function of 
the distance, $L$, starting from the point of accretion 
\begin{equation}
\Delta M(L)\simeq 2.4\times 10^{-12} {\rm TeV} \,(L+L_0)^{-1/2}  \frac{M_{\rm min}^2}{M_{P}^{3/2}},
\end{equation}
where $L$ is measured in meters and
\begin{equation}
L_0=2.3 10^{-17}{\rm m}  \,\frac{M_{\rm min}^4}{M_{P}^3}
\end{equation}
Fig.1 shows the dependence of the average energy emitted by the black hole per meter on $L$ at 
five different choices of $M=M_{\rm min}^4/M_P^3$. The initial excess energy has been chosen to 
be $\Delta M=0.5$ GeV.
\begin{figure}[htbp]
\begin{center}
\includegraphics[width=4in]{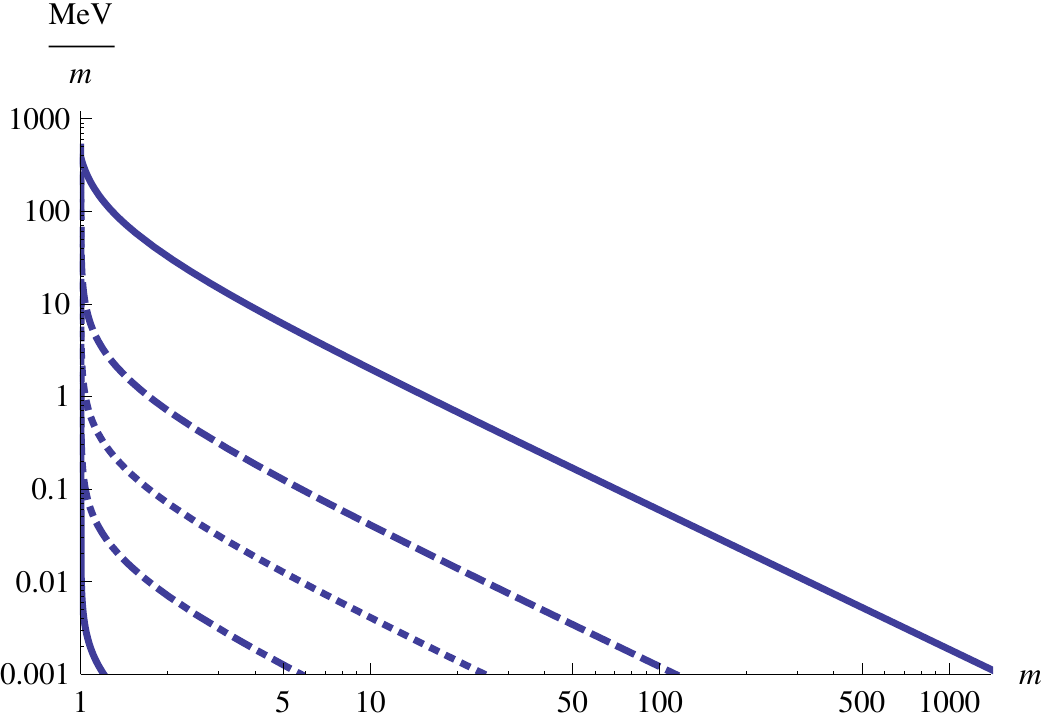}
\caption{The average deposited energy per meter for $M=M_P^{(4)} $ (solid line) , $M=10^{12}$ TeV 
(dashed), $M=10^{10}$ TeV (dotted), $M=10^{8}$ TeV (dash-dotted),$M=10^{5}$ TeV (bottom solid line)}
\label{ionization}
\end{center}
\end{figure}

We can conclude from Fig.1. that if $M=M_P^{(4)}=2.4\times 10^{15}$ TeV there is a significant chance 
of observing a series of emissions of $\gamma$ rays and energetic electrons at large distances from 
the point of accretion.  At lower values of $M$, certainly for $M<10^{10}$ TeV most of the energy is 
deposited within a very short distance and only some hard X-ray photons can be emitted up to 
approximate distance of 10 m from the point of accretion. 

The spatial resolution of the Super-Kamiokande detector for  sub-GeV particles is around 0.3 m 
\cite{kamiokande}. We can see that at $M\leq10^5$ TeV the decay  appears  to be completely localized 
at the point of accretion.

The Super-Kamiokande detector is eminently suitable for discovering stable microscopic black holes, 
provided the abundance of black holes in dark matter is large enough and $M_P$ is small enough. 
The approximate limits on these quantities are shown in Fig. 2. along with the upper bound on the flux 
(\ref{upper1}) and (\ref{upper2}). A further limitation on the mass of stable black holes comes from 
accelerator experiments at Fermilab or the LHC.  All these limits are plotted in Fig. 2.
\begin{figure}[htbp]
\begin{center}
\includegraphics[width=4in]{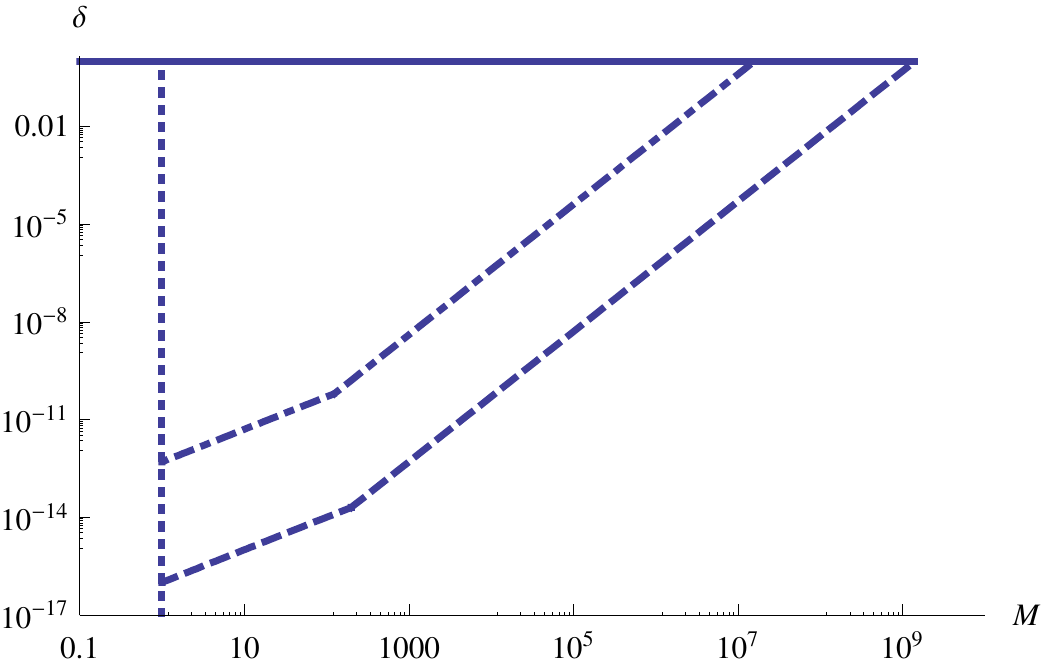}
\caption{The $\delta$-$M$ plane: (1) The region left of the dotted line is excluded on the basis of 
accelerator experiments, (2) The region above the dash-dotted line is excluded by limits on the flux in 
proton decay experiments, (3) The region between the dash-dotted and dashed lines represents stable 
microscopic black holes that are amenable to discovery at Super-Kamiokande, (4) In the region 
under the dashed line  the event rate is too low for discovery with currently available detectors, (5) 
the $\delta>1$ region above the solid line is excluded because of astronomical limits on dark matter.} 
\label{admissible}
\end{center}
\end{figure}

One may consider the scenario of an eleven dimensional theory, in which there is a lower bound on the 
black hole mass in the presence of the Lovelock term $L_5$, compactified above the Planck scale of 
$M_P\lesssim 10^{15}$ TeV. Then, if the minimal mass is of the same order of magnitude, all the bounds 
are satisfied, even if all of dark matter is constituted from black holes.  Unfortunately, 
using (\ref{eventrate}),  if $M_{\rm min}\sim M_P\gtrsim 10^9$ TeV the event rate at Super-Kamiokande 
drops below 1/year and observing an accretion and decay of a black hole becomes not feasible.  
One may significantly increase this upper limit  if one also searches for series of $\gamma$ emissions 
that a black hole emits after it accreted a parton in the rock outside the fiducial volume of the detector, 
as shown by Fig. 1.

  \section{ Relativistic black holes in a medium}

In low scale gravity, $2.4\times 10^{15}\,{\rm TeV}=M_P^{(4)}\geq M_P\geq 1$ TeV,  high energy 
particles can create black holes.  If the primary energy is $E>> M_P^2/m_p$, such as may be in energetic 
cosmic rays, then the produced black holes have relativistic speed and they may or may not slow down 
and evaporate while passing through matter. The mass and speed of the black hole changes due to 
the absorption of matter particles and Hawking radiation.  We will investigate below what are the 
conditions for the black hole, undergoing multiple collisions, to slow down and evaporate before 
accreting all matter available to it \cite{argyres}.

Since in a brane-world scenario all particles save the graviton and Kaluza-Klein modes are confined to 
the brane and the Kaluza-Klein modes contribute little \cite{emparan}, despite the fact that the black 
holes are higher dimensional, we write the equation for the time dependence of the mass in the rest 
system of the black hole using 4-dimensional quantities. The change of mass of the black hole as 
described in the rest system of the black hole is 
\begin{equation}
\frac{dM}{dt}=c\sigma \rho \gamma^2v-g\frac{\pi^2}{120}4\pi r_h^2T_H^4,
\label{time}
\end{equation}
where $g=\sum_s \Gamma_s g_{\rm eff}^s$, where $\Gamma_s$ and $g_{\rm eff}^s$ are the gray body 
factor and effective number of degrees of freedom for particle species $s$ \cite{chamblin}. 
$c$ is the effectiveness of turning the energy of a parton into the mass of the black hole.   Stefan's constant 
per massless degree of freedom is $\pi^2/120$,  $\rho$ is the mass density of the medium and $\gamma$ 
is the time dependent Lorentz factor. $r_h$ is the radius of horizon. The multiplier $\gamma^2$ should be 
included in the absorption term because on one hand the black hole sees a Lorentz contracted 
distribution and the other hand the energy of each particle of mass $m$ absorbed is $m\gamma$.  
Another way of stating this is that the accretion rate in the star's rest system is
\begin{equation}
\left.\frac{dM}{dt}\right|_a=c\sigma\, \rho\, \Delta v,
\end{equation}
where $\Delta v$ is the relative speed.  Now transforming into the system of the black hole while 
$t\rho\to \gamma^2t\rho$ and the relative speed is unchanged. We defined
\begin{equation}
r_h=\sqrt{\mu^2-2\alpha} .
\end{equation}
where $\mu$, the radius of horizon, is related to the mass as in (\ref{mass}). $T_H$ is the Hawking 
temperature,
\begin{equation}
T_H=\frac{ \sqrt{\mu^2-2\alpha} }{2\pi(\mu^2+2\alpha)}.
\end{equation}

Since $\gamma$ is also time dependent we need another equation to find both $\mu$ and $g$. We can 
readily obtain such an equation if we consider the process in the rest system of the medium. Then using 
momentum conservation in the rest frame of the star and assuming that there is no momentum transfer to 
matter not absorbed by the black hole we obtain
\begin{equation}
\frac{d(M\gamma v)}{dt}=-g\gamma v\frac{\pi^2}{120}4\pi r_h^2T_H^4,
\label{energy}
\end{equation}
where we assumed that the c.m.s of evaporating particles moves with the same speed as the black hole. 

Combining (\ref{time}) and (\ref{energy}) we obtain an equation for the Lorentz factor
\begin{equation}
\frac{d\gamma}{dt}=-\frac{\sigma}{M} c\rho (\gamma^2-1)^{3/2}
\label{gamma}
\end{equation}
In the Schwarzschild limit ($M>>\alpha M_P^3$, where $M_P$ is the Planck mass) the ratio 
\begin{equation}
\frac{\sigma}{M}=\frac{4}{3M_P^3}
\end{equation}
is a constant and the equation for $\gamma$ decouples.
Introducing the notation $x=4 c\rho/(3 M_P^3)$ and using the initial condition, $\gamma=\gamma_0$ at 
$t=0$, the solution of (\ref{gamma}) is 
\begin{equation}
v=\frac{v^0}{1+x\, t \,v_0}
\end{equation}

$\mu(y)$ can be obtained as a functional of  $\gamma$ from (\ref{time})  in the Schwarzschild approximation, 
with initial condition $\mu^2(0)=\mu_0^2=M_0\,/\,(3\pi M_P^3)$,
\begin{equation}
\mu^2(y)=h(y)\sqrt{\mu_0^4-2\frac{\beta}{x}\int_0^y[h(z)]^{-2}dz}
\label{massequation}
\end{equation}
where
$\beta=g\pi^{-2}M_P^{-3}/180$, $y=x\,t$, and
\begin{equation}
h(y)=\exp\left\{\int_0^y[\gamma(z)]^2dz\right\}=\gamma_0\sqrt{(y v_0+1)^2
-v_0^2}.
\end{equation}
Calculating the integrals in (\ref{massequation}) we get 
\begin{equation}
\mu^2=\gamma_0\left[(x\,t\, v_0+1)^2-v_0^2\right]^{1/2}\left[\mu_0^4-\frac{\beta}{\gamma_0^2\,x\,v_0}\log\left(\frac{(1+v_0)(1+v_0x\,t-v_0)}{(1-v_0)(1+v_0x\,t+v_0)}\right)\right]^{1/2}
\label{musquared}
\end{equation}

Now there are two possible scenarios. The first one is when  $\mu^2\to \infty$ when $t\to \infty$, if
\begin{equation}
\mu_0^4>\frac{\beta}{\gamma_0^2 \,x\,v_0}\log\left(\frac{1+v_0}{1-v_0}\right)\simeq\frac{\beta}{\gamma_0 ^2\,x\,v_0}\log(4\gamma_0^2).
\end{equation}
In that case the mass of the black hole grows indefinitely, i.e it gobbles up all matter available. In the 
opposite case an Einstein black hole evaporates completely in finite time.  The evaporation time can be 
obtained from finding the zero of the last factor of (\ref{musquared}). For a Lovelock black hole, which 
approaches  a finite mass as $t\to\infty$,  we must slightly modify (\ref{musquared}) at large times.  
 The difference between Einstein and Lovelock black holes in no way modifies our conclusions concerning 
the fate of cosmic ray black holes in neutron stars. 
 
Using $M_0\gamma_0=E_0$, the primary energy, we obtain the following upper bound on the primary energy 
for the black hole to evaporate already in the crust of the neutron star (with density $10^9$ kg/m$^3$)
\begin{equation}
E_0<10^{11}{\rm TeV}\, M_P^3.
\end{equation}

According to current theory and experiment the cosmic ray spectrum cuts off at  $E_{\rm max}=5\times 10^{7}$ 
TeV due to the Greisen-Kuzmin-Zatsepin effect \cite{greisen}.  We estimate that a black hole created by a 
cosmic ray proton of $E<E_{\rm max}$ in the crust of a neutron star will lose its energy in approximately 
$l\simeq 1/x\simeq$ 1$\mu$m.

\section{Conclusions}

In more than 4 space-time dimensions Lovelock terms make a non-trivial contribution to the equations of 
motion. Putting it in stronger terms, there is no reason known to us why Lovelock terms should not appear 
in the gravitational action.  Their presence significantly modifies the physics of small black holes.  Some of 
these modifications, especially in odd dimensions, in which black holes have a lower bound on their mass, 
have been investigated in the past \cite{rizzo} \cite{rizzo1} \cite{alexeyev}.  At the lower bound odd 
dimensional Lovelock black holes, just like extremal black holes, have vanishing Hawking temperature  
\cite{myers}\cite{wiltshire} \cite{dehghani}. At least semi-classically, at the end of their decay process, 
these black holes become stable in vacuum. 

If D is even, then after the compactification of an odd number of dimensions, we believe that small odd 
dimensional black holes can be caged without a substantial change of their metric and thus they retain 
their property of having a minimal mass and vanishing Hawking temperature. This could be ascertained 
in an expansion in the inverse of the compactification radius \cite{harmark} \cite{karasik}.  Alternatively, 
in even dimensions, if the compactification radius of an odd number of dimensions is smaller than 
$\mu^{(D)}$ black holes become black branes. Such black branes retain the combination of properties 
of having a minimal mass and vanishing Hawking temperature at the minimal mass. This will be the 
subject of a future publication.  

If black holes become stable upon reaching their minimal mass then their number density can be reduced 
only by collisions of black holes. Then by necessity, near minimum mass black holes form  a part of dark 
matter. One major concern of this paper has been to find upper limits on the contribution of stable 
microscopic (of mass between 1 TeV and the four dimensional Planck mass) black holes to dark matter.  
We have provided an approximate upper bound on their  abundance using existing data at 
underground experiments.  In fact, we also show that using the extreme assumption that all of dark 
matter is constituted from stable microscopic black holes their mass is at least $10^4$ TeV, but probably 
more than $10^7$ TeV.  In doing so we show that stable black holes traversing through underground 
chambers have a very peculiar characteristic signals, especially if their minimum mass is larger then 
$10^5$ TeV.  In the range $M_{\rm min}<10^5$ TeV their decay is local, often imitating a low energy 
($E\lesssim 0.5$ GeV) neutrino interaction.  However, at higher energies, especially at $M_{\rm min}>10^{10}$ 
TeV, after a localized partial evaporation they continue emitting lower and lower energy particles 
(mostly photons and possibly electrons) along a track as they progress through the detector.  
Unfortunately, if $M_{\rm min}\sim M_P\gtrsim10^{10}$ TeV then the event rate at the Super-Kamiokande 
detector becomes so low that the observation the black hole evaporation process becomes very difficult. 

Black holes that form part of dark matter accumulate in a central region of neutron stars.  There they 
annihilate by colliding with each other.  Eventually they reach equilibrium with the incoming flux and 
then they produce heat though possibly at a lower rate than other components of  dark matter 
\cite{kouvaris}. In ordinary stars, however the surface gravity is not sufficient to keep the black holes 
from escaping unless the $M_{\rm min}>$100 TeV, in which case the average velocity of black holes 
becomes smaller than the escape velocity.  However, if the $M_{\rm min}\sim M_P\gtrsim10^{15}$ TeV 
the mean free path in black holes and in ordinary stars becomes comparable with the size of the star 
and most of the black holes are not captured. 

Among other aspects of the physics  of black holes we investigated the accretion of relativistic 
(cosmic ray) black holes in matter.  We find that all black holes created by cosmic ray protons below 
the Greisen-Kuzmin-Zatsepin limit \cite{greisen} slow down and lose mass in stars, among others, in 
neutron stars.

Finally, we return to a problem, which we already mentioned in the Introduction. Two further steps 
are required to complete our scenario, either of which could potentially destabilize minimum  mass 
black holes.  Suppose that $D$ is the (odd) number of dimensions of Lovelock gravity, which we 
have considered in this paper.  Since there are only 4 non-compact dimensions in our world we 
still need to compactify $D-4$ dimensions.  

The second problem concerns the existence of dimensions beyond $D$. The total number of dimensions 
of the world, ${\mathfrak{D}}\geq D$. $\mathfrak{D}$ may be even or odd. We must embed black 
holes into the $\mathfrak{D}$ dimensional space. In the next paragraph we will sketch a possible 
solution to these problems.

Suppose the compactification scales of $\mathfrak{D}-D$ and $D-4$ dimensions are 
$\mathfrak{L}$ and $L$, where $\mathfrak{L}<<L $. We assume that  $D$ is odd and all Lovelock 
terms up to $n=(D-1)/2$ order appear in the action.  The minimum mass should be small enough, 
such that black holes close to the minimum mass are caged in the $D-4$ compact dimensions, 
but at the same time it should be large enough so that black holes are extended as black branes 
into the extra $\mathfrak{D}-D$ dimensions.  In other words, the condition for the existence of 
a stable minimum mass is that $\mu^{(D)}$ defined in (\ref{mass})  is in between the two scales, 
i.e  $c_\mathfrak{L}\,\mathfrak{L} <\mu^{(D)}<c_L\, L $, where $c_\mathfrak{L}$ and $c_L$ are 
$O(1)$ dimensionless constants. $c_\mathfrak{L}$ is defined to be large enough to avoid the 
Gregory-Laflamme instability \cite{gregory}, while $c_L$ is chosen to be small enough to have the 
black hole safely caged in $D-4$ dimensions.   The details of this scenario are left to a future publication.

\section{}
\begin{acknowledgments}
This work is supported in part by the DOE grant \# DE-FG02-84ER40153. The authors are indebted to 
Philip Argyres for informing them about his study of the accretion of black holes, produced by 
cosmic rays, in neutron stars and for valuable discussions. L.C.R.W. thanks  S.Chivukula,  F. P. Esposito, 
P.Frampton,  C.Kouvaris,  R.Schnee , F.Sannino for discussions  and CP3 Origins Institute  in Odense  
and Aspen Center for Physics for their hospitality.
\end{acknowledgments}


\begin{thebibliography}{10} 
\bibitem{ADD}N.Arkani-Hamed, S.Dimopolous and G.Dvali, Phys.Lett.B429:263-272,1998 :hep-ph/9803315;I. Antoniadis,  , N.Arkani-Hamed  , S.Dimopoulos,   , G.R. Dvali,    Phys.Lett.B436:257-263,1998.
e-Print: hep-ph/9804398
\bibitem{RS1} L.Randall and R.Sundrum, , Phys.Rev.Lett. 83, 3370, (1999).
\bibitem{RS2}  L.Randall and R.Sundrum, , Phys.Rev.Lett. 83, 4690 (1999).
\bibitem{dimopoulos} S. Dimopoulos and G. Landsberg, PRL {\bf 87}, 161602 (2001)
\bibitem{cosmic} L.A. Anchordoqui, J.L. Feng, H. Goldberg and A.D. Shapere, Phys. Rev. D65:124027 (2002) hep-ph/0112247
\bibitem{production}S. Giddings and E. Katz J. Math. Phys. 42:3082-3102 (2001); T. Banks and W. Fischler, Preprint hep-th/9906038; S. 
Dimopoulos and G. Landsberg, Phys. Rev. Lett. 87:161602 (2001); S.B. Giddings and S. Thomas, Phys. Rev. D65:056010.
(2002); D.M. Eardley and S.B. Giddings, Phys. Rev. D66:044011 (2002) gr-qc/0201034.
\bibitem{giddings}
Steven B. Giddings, (UC, Santa Barbara) , Michelangelo L. Mangano,
Published in Phys.Rev.D78:035009,2008
e-Print: arXiv:0806.3381 [hep-ph]
\bibitem{gregory} R. Gregory and R. Laflamme, Phys.Rev.Lett. 70, 2837 (1993),	arXiv:hep-th/9301052v2
\bibitem{lovelock} C. Lanczos, Ann. Math, {\bf 39}, 842 (1938), D. Lovelock, J. Math. Phys. {\bf 12}, 498 (1971)
\bibitem{blackbrane} C. Sahabandu, P. Suranyi, C.Vaz, and L.C.R. Wijewrdhana, Phys. Rev. D{\bf 73}, 044009 (2006).
\bibitem{rizzo1} T. Rizzo, JHEP 0506:079,2005. 
e-Print: hep-ph/0503163
\bibitem{rizzo} T. Rizzo, Class.Quant.Grav.23:4263-4280,2006. 
\bibitem{alexeyev} S.O. Alexeyev, A. Barrau, G. Boudoul, O. Khovanskaya, and M. Sazhin, Class. Quant. Grav. {\bf 19}, 4431 (2002)
\bibitem{heterotic}B. Zwiebach, Phys.Lett. B156 (1985) 315.
 B. Zumino, Phys. Rept. 137 (1986) 109, D J. Gross and  E. Witten,  Nucl.Phys.B277:1,1986. 
 E.Bergshoeff and M. de Roo,Nuc.Phys.B328(1989)939: M.C.Bento and O.Bertolami, Phys.LettB.228(1989)348 and Phys.Lett.B,368(1996)198.
\bibitem{boulware}David G. Boulware and Stanley Deser
Phys.Rev.Lett.55:2656,1985.
\bibitem{dama} R. Barnabei et al. European Physical Journal C 56: 333. (2008) arXiv:0804.2741 [astro-phys].
\bibitem{cdms}  Z. Ahmed et. al. Science, 327, 1619 (2010).
\bibitem{frampton} 
Paul H. Frampton, 
Published in JCAP 0910:016,2009. 
e-Print: arXiv:0905.3632 [hep-th];P. Frampton,, Masahiro Kawasaki, Fuminobu Takahashi, Tsutomu T. Yanagida, . 
Published in JCAP 1004:023,2010. 
\bibitem{kouvaris} C.Kouvaris, Peter Tinyakov, 
e-Print: arXiv:1004.0586, 
Arnaud de Lavallaz, Malcolm Fairbairn, 
e-Print: arXiv:1004.0629
 C.Kouvaris, 
Published in Phys.Rev.D77:023006,2008. 
e-Print: arXiv:0708.2362.
\bibitem{bekenstein} J.D. Bekenstein, Phys. Rev. D7, 2333 (1973).
\bibitem{databook} W.M. Yao et.al. [Particle Data Book], "Review of Particle Properties," J. Phys. G {\bf 33}, 1 (2006).
\bibitem{wiltshire}J.Wiltshire, Phys.Rev.D38:2445,1988 .
\bibitem{myers}
R. C. Myers, J. Z. Simon, Phys.Rev.D38:2434-2444,1988.
\bibitem{dehghani}
M.H. Dehghani, R. Pourhasan,
Published in Phys.Rev.D79:064015,2009.
e-Print: arXiv:0903.4260 [gr-qc]
\bibitem{harmark} Troels Harmark, Niels A. Obers, (Bohr Inst.) . Sep 2003. 26pp. 
Published in Nucl.Phys.B684:183-208,2004; Hideaki Kudoh, (Kyoto U.) , Toby Wiseman, 
Published in Prog.Theor.Phys.111:475-507,2004. 
e-Print: hep-th/0310104; Barak Kol, 
Published in JHEP 0510:049,2005. 
e-Print: hep-th/0206220
\bibitem{karasik} D. Karasik, C. Sahabandu, P. Suranyi, L.C.R. Wijewardhana, Published in Phys.Rev.D71:024024,2005. 
e-Print: hep-th/0410078
\bibitem{hawking} Don N. Page, S.W. Hawking, 
 Astrophys. J. 206:1-7,1976.
 \bibitem{emparan} R. Emparan, G,. Horowitz, and R.C. Myers, PRL {\bf 85}, 499 (2000).
 \bibitem{anchordoqui} L. Anchordoqui and H. Goldberg,  Phys. Rev. D67: 064010 (2003), hep-ph/0209337v4 
 \bibitem{lower} U. Mahanta and Anindya Datta, Phys.Lett. {\bf B483}, 196 (2000).
\bibitem{unruh} W.G. Unruh, Phys. Rev. {\bf 14}, 3255 (1976).
 \bibitem{graviton} V. Cardoso, M. Cavaglia, L. Gualtieri, arXiv:hep-th/0512002 published at Phys.Rev.Lett. 96 (2006) 071301.
 \bibitem{minnesota} R.M. Bionta et. al. Phys. Rev. Lett. 51, 2717 (1983).
 \bibitem{kamiokande} "Atmospheric neutrino oscillation analysis with solar terms in 
Super-Kamiokande," Yumiko Takenaga, PhD theses, University of Tokyo, February 15, 2008
\bibitem{argyres} P. Argyres completed such a calculation and reached conclusions, similar to ours. 
\bibitem{greisen} K. Greisen, Phys. Rev. Lett. 16, 748170 (1966); 
G. T. Zatsepin and V. A. Kuz'min, 
Jour. Exp. Theor. Phys. Lett.,  4, 78 (1966)
\bibitem{chamblin}A. Chamblin, F. Cooper, and G. C. Nayak, Phys. Rev. {\bf D69} 065010 (2004).
\end{thebibliography}
\end{document}